\documentclass[
aps,
prd,
preprint,
superscriptaddress,
amsfonts,
amssymb,
amsmath,
preprintnumbers,
floatfix,
nofootinbib,
unsortedaddress,
showpacs,
tightenlines,
floats,
a4paper,
]{revtex4}
\usepackage{graphicx}
\usepackage{slashbox}
\usepackage{float}
\usepackage{amssymb,amsmath}
\begin{document}
\title{Self-consistent initial conditions for \\
primordial black hole formation}
\author{A.~G.~Polnarev}

\affiliation{Astronomy Unit, School of Physics and Astronomy, \\
Queen Mary
University of London, \\ Mile End Road, London E1 4NS, United Kingdom
}

\author{Tomohiro Nakama}

\affiliation{Department of Physics,
Graduate School of Science,\\ The University of Tokyo, Bunkyo-ku,
Tokyo 113-0033, Japan
}

\affiliation{Research Center for the Early Universe (RESCEU),\\
Graduate School of Science, The University of Tokyo, \\ Bunkyo-ku,
Tokyo 113-0033, Japan
}

\author{Jun'ichi Yokoyama}

\affiliation{Research Center for the Early Universe (RESCEU),\\
Graduate School of Science, The University of Tokyo, \\ Bunkyo-ku,
Tokyo 113-0033, Japan
}

\affiliation{Kavli Institute for the Physics and Mathematics 
of the Universe,\\
The University of Tokyo, Kashiwa, Chiba 277-8568, Japan}

\date{\today}
\def\ve{\varepsilon}
\def\al{\alpha}
\def\be{\begin{equation}}
\def\ee{\end{equation}}
\def\ep{\epsilon}
\def\pa{\partial}
\def\ti{\tilde}
\def\fr{\frac}
\def\de{\delta}
\def\ga{\gamma}
\def\de{\delta}
%\begin{titlepage}
\preprint{RESCEU-10/12}

%\vskip 1cm

\date{\today}

\small

\begin{abstract}

For an arbitrarily strong, spherically symmetric super-horizon curvature 
perturbation, we present analytic solutions of the Einstein equations
in terms of the asymptotic expansion over the ratio of the  Hubble 
radius to the length-scale of the curvature perturbation to set initial conditions 
for numerical computations of primordial black hole formation. To obtain this solution we develop a recursive method 
of quasi-linearization which reduces the problem to a system of 
coupled ordinary differential equations for the $N$th 
order terms in the asymptotic expansion with sources consisting of a
non-linear combination of the lower order terms. 
%For an arbitrary precision requirement, determined 
%by the intended accuracy and stability of the computer code, our analytical 
%solution yields the optimal truncated asymptotic expansion (up to \underline{seventh} order) 
%which can be used to 
%find the upper limit on when the initial conditions 
%expressed in terms of such a truncated expansion should be set. Examples 
%of how 
%these truncated solutions provide initial 
%conditions with given accuracy for different radial profiles of 
%curvature perturbations are presented.

\end{abstract}

%\pacs{98.80.Cq}

\maketitle
%\end{titlepage}

\section{Introduction}
%%%%%%%%%%%%%%%%%%%%%%%%%%%%%%%%%%%%%%%%%%%%%%%%%%%%%%%%%%%%%%%%%%%%%%%%%%%%%
The idea that large-amplitude matter overdensities in the
Universe could have collapsed through self-gravity to form
primordial black holes (PBHs) was first put forward by
Zel'dovich and Novikov \cite{Zel'dovich-1974}, and then independently by
Hawking \cite{Hawking:1971ei}, more than three decades ago. This theory
suggests that large-amplitude inhomogeneities in the very
early universe overcome internal pressure forces and collapse
to form black holes. A lower threshold for the amplitude
of such inhomogeneities was first
provided by Carr \cite{Carr:1974nx,Carr:1975qj} for a 
radiation-dominated epoch. 
The PBH
contribution to the energy density increases with time
during this epoch. For this reason,
the PBHs formed considerably before the end of radiation-domination, possibly even 
before radiation-domination \cite{1980PhLB...97..383K}, affect various cosmological and astrophysical
processes even if their initial abundance is tiny.
PBHs with mass smaller than $\sim 10^{15}$g would have evaporated
through Hawking radiation \cite{Hawking:1974rv} and their abundances
are constrained by big-bang nucleosynthesis \cite{Zel'dovich:1977sta,Novikov:1979pol,
1978AZh....55..231V,1978PAZh....4..344V,Miyama:1978mp,Kohri:1999ex} and the gamma-ray
background \cite{Page:1976wx,MacGibbon:1987my,MacGibbon:1991vc}, 
while holes with larger masses are constrained by dynamical and lensing effects \cite{Paczynski:1985jf} 
and by the stochastic gravitational wave background
\cite{Saito:2008jc,Saito:2009jt}.  
All these constraints are updated and summarized in
\cite{Carr:2009jm}. 

Since the probability of PBH formation depends 
crucially on the statistical characteristics of the random field of primordial 
perturbations,
%probability of
PBHs provide a useful and unique tool to obtain independent
constrains on the primordial power spectrum %mean amplitude
of inhomogeneities on extremely small scales
which cannot be probed by any other methods.
To make this cosmological tool more reliable, we must improve the prescription of the initial 
conditions \cite{Polnarev:2006aa}. Self-consistent initial 
conditions are very important for  calculations of the probability
of PBH formation \cite{Hidalgo:2008mv}  and for relativistic 
hydrodynamical  computations \cite{1978SvA....22..129N,1980SvA....24..147N,
1979ApJ...232..670B,Shibata:1999zs}. 
According to such computations the 
pressure gradients in the collapsing configuration play an extremely 
important role and are directly determined by 
the curvature profile in the initial configuration.
 
Since PBHs can form only 
from highly non-linear curvature perturbations, 
the
 initial conditions of their formation must be consistent with the underlying non-linear theory such as
the general relativity.
%and according 
%to  non-linear theory such as General Relativity,  not all initial 
%conditions are compatible with Einstein equations, i.e. are not 
%self-consistent,  such dangerous possibility is unavoidable. 
The main objective of the present paper is to 
exclude the possibility that even highly sophisticated computer 
simulations could produce irrelevant results due to
inconsistency of the initial conditions.
For example, if we assume that the 
Universe outside the configuration is spatially flat, the mass 
of a perturbed configuration of radius $r$ should be equal to the 
mass of unperturbed sphere of the same radius. As a result, a 
self-consistent density profile should be non-monotonic, {\it i.e.} along 
with the region of density excess, it should contain a region 
of density deficit, which drastically changes the effect of 
pressure gradients \cite{1978SvA....22..129N}.

For an arbitrarily strong spherically symmetric super-horizon 
curvature perturbation, we present an analytic solution of the 
Einstein equations
in terms of an asymptotic expansion over the ratio of the  
Hubble radius to the length-scale of the curvature perturbation 
under consideration. This method is similar to the gradient expansion 
\cite{Adv.Phys.12.185,CQG.9.1943,PhysRevD.49.2759,0264-9381-13-4-010,0264-9381-19-14-322,PTP.117.633} 
(previously known as the anti-Newtonian expansion \cite{PTP.54.730}) in spirit, but thanks to 
the spherical symmetry we can construct a solution to arbitrary higher order in this ratio from a 
single function characterising the curvature profile. Note that the lowest-order solution in this program 
has been obtained in \cite{1978SvA....22..129N,Polnarev:2006aa,Musco:2008hv}.
%In contrast to the previous papers 
%\cite{1978SvA....22..129N,Polnarev:2006aa,Musco:2008hv},
%where only the first two terms of  
%such an expansion were considered, in the present paper 
Here we develop 
the recursive method of quasi-linearization which reduces the 
problem to a  system of coupled ordinary differential equations 
for terms of $n$th order in the asymptotic expansion of the 
density, pressure, velocity and metric, with sources which contain 
a non-linear combination of the lower order terms. Using this method, 
we obtain analytic expressions for all terms.

The curvature profile, $K_\mathrm{i}(r)$, to be defined below,
 appears in the 
 source terms on the 
right-hand side of the relevant equations and %we can say that 
 deviations from homogeneity in
density and velocity are generated 
by the inhomogeneity of the curvature. To avoid confusion, we should 
say that these deviations do not involve small cosmological 
perturbations at all. Statistical characteristics of small
 perturbations  are only relevant in this context  when one 
calculates the probability of finding a configuration with a high 
amplitude perturbation of the metric.  

Dropping all terms of order 
greater than $N$, we obtain truncated asymptotic solutions of $N$th order. 
Then for the arbitrary precision required by the intended 
accuracy and stability of the computer code, 
we obtain  an upper limit on the time when 
such an $N$th order truncated expansion can be used
to set the initial conditions of fully non-linear numerical simulations of 
PBH formation. Later initial 
times obviously correspond to 
shorter computer runs. Thus our analytic solution helps to 
optimize numerical computations. 

The rest of the paper is organized as follows. In \S I\hspace{-.1em}I basic equations are derived and in \S I\hspace{-.1em}I\hspace{-.1em}I
expansion coefficients are defined and their properties are described. Then in \S I\hspace{-.1em}V we derive 
the recursive formulae 
for the coefficients in our problem 
%In \S V the relation between the required precision for the truncated asymptotic expansion and 
%the upper limit on the moment of time is discussed. 
and they are solved in \S V for several specific initial curvature profiles. 
\S V\hspace{-.1em}I 
%\S V\hspace{-.1em}I\hspace{-.1em}I 
is devoted to
discussion and drawing conclusions.

\section{Mathematical formulation of the problem}
\subsection{The Misner-Sharp equations}
Assuming spherical symmetry, it is convenient to divide the collapsing matter
into a system of concentric spherical shells and to label each shell with a
Lagrangian comoving radial coordinate $r$. Then the metric can
be written in the form used by Misner and Sharp \cite{Misner:1964je}:
\begin{equation}
ds^2=-a^2dt^2+b^2dr^2+R^2(d\theta^2+\sin^2\theta d\phi^2),\label{1}
\end{equation}
where $R$, $a$ and $b$ are functions of 
$r$ and the time coordinate $t$. We consider a perfect fluid with energy density $\rho(r,t)$ and pressure $p(r,t)$ 
and constant equation-of-state parameter $\ga$, $p(r,t)=\ga \rho(r,t)$.
Expressing the proper time derivative of $R$ as
\begin{equation}
u\equiv \frac{\dot{R}}{a},\label{2}
\end{equation}
with a dot denoting a derivative with respect to $t$, we derive equations of motion for these variables as follows.

First, from the $(^0_r)$ component of the Einstein equations, we find

\begin{equation}
\frac{\dot{b}}{b}=\frac{au'}{R'},\label{3}
\end{equation}
while the Euler equation yields
\begin{equation}
\frac{a'}{a}=-\frac{\gamma}{1+\gamma}\frac{\rho'}{\rho},\label{4}
\end{equation}
where a prime denotes differentiation with respect to $r$. We define the mass within the shell of proper radius $R$ by
\be
M(r,t)=4\pi\int^{R(r,t)}_0\rho(r,t)R^2dR,\label{defofM}
\ee
which gives
\begin{equation}
M'=4\pi \rho R^2R'.\label{5}
\end{equation}
Using (\ref{3}), the $(^0_0)$ component of the Einstein equations becomes
\begin{equation}
\frac{R'^2}{b^2}=1+u^2-\frac{2GM}{R},\label{7}
\end{equation}
and (\ref{defofM}) can then be expressed as
\be
M(r,t)=\int^r_0\rho\left(1+u^2-\fr{2GM}{R}\right)^\fr{1}{2}dV,\label{M}
\ee
where $dV\equiv 4\pi R^2bdr$ is the proper volume element. Equation (\ref{M}) shows that $M$ includes
contributions from both the kinetic energy and the gravitational potential energy.
Finally, combining (\ref{3})$\sim$(\ref{7}), the evolution equation of $M$ becomes
\begin{equation}
\dot{M}=-4\pi p R^2\dot{R}.\label{6}
\end{equation}
\subsection{
Quasi-homogenous asymptotic equations in new variables }
We consider the evolution of a perturbed region described by the above equations embedded in a 
flat Friedmann-Lemaitre-Robertson-Walker (FLRW) Universe with metric
\be
ds^2=-dt^2+S^2(t)(dr^2+r^2d\theta^2+r^2\sin^2\theta d\phi),
\ee
which is a particular case of (\ref{1}). The scale factor in this background evolves as
\begin{equation}
S(t)=\left(\fr{t}{t_\mathrm{i}}\right)^{\alpha} ,\quad\alpha\equiv\frac{2}{3(1+\gamma)},
\end{equation}
where $t_\mathrm{i}$ is some reference time.

We denote the background solution with a suffix 0.
In terms of the metric variables defined in (\ref{1}), we find
\be
a_0=1,\: b_0=S(t),\:  R_0=rS(t).
\ee
The background Hubble parameter is
\begin{equation}
H_0(t)=\fr{\dot{R_0}}{a_0R_0}=\frac{\dot{S}}{S}=\frac{\alpha}{t},
\end{equation}
and the energy density is calculated from the Friedmann equation,
\begin{equation}
\rho_0(t)=\frac{3\alpha^2}{8\pi Gt^2}.
\end{equation}
We introduce a variable $H$ defined by
\begin{equation}
H(t,r)\equiv \fr{\dot{R}}{aR}=\fr{u}{R}
\end{equation}
and another new variable $\tilde{H}$ by rewriting $H$ as
\begin{equation}
H(t,r)=H_0(t)\tilde{H}(t,r).
\end{equation}
The tilde-variable $\tilde{H}$, as well as most of the other tilde-variables introduced
below, represent deviations of the solutions from the corresponding ones in the
flat FLRW universe. Specifically we find
\be
a(t,r)=a_0(t)\tilde{a}(t,r)=\ti{a}(t,r),
\ee
\be
b(t,r)=b_0(t)\ti{b}(t,r)=S(t)\ti{b}(t,r),
\ee
\be
R(t,r)=R_0(t)\ti{R}(t,r)=rS(t)\ti{R}(t,r),
\ee
\be
\rho(t,r)=\rho_0(t)\ti{\rho}(t,r)\propto t^{-4\al}\ti{\rho}.
\ee
We define another variable $\tilde{\mu}$ by
\begin{equation}
M=\frac{4\pi}{3}\rho_0R^3\tilde{\mu},\label{17}
\end{equation}
and the curvature profile $K(t,r)$ is defined by rewriting $b$ as
\be
b(t,r)=\frac{R'(t,r)}{\sqrt{1-K(t,r)r^2}}\label{16}.
\ee
$K(t,r)$ vanishes outside the perturbed region so that the solution asymptotically approaches the background FLRW
 solution at spatial infinity.

We denote the comoving radius of a perturbed region by $r_\mathrm{i}$, whose
precise definition will be given later, 
 and define a dimensionless parameter
$\ep$ in terms of the square ratio of the Hubble radius  $H_0^{-1}$ to the physical
length scale of the configuration,
\be
\ep \equiv \left(\frac{H_0^{-1}}{S(t)r_\mathrm{i}}\right)^2
=(\dot{S}r_\mathrm{i})^{-2}=\frac{t_\mathrm{i}^{2\al}t^{\beta}}{\al^2 r_\mathrm{i}^2},\quad \beta\equiv 2(1-\al).
\ee
When we set the initial conditions for PBH
formation, the size of the perturbed 
region is much larger than the
Hubble horizon. 
This remains the case until the horizon mass becomes larger than
the PBH mass. The horizon mass grows with cosmic time
(for the radiation-dominated regime,
the growth is directly proportional to time).
This means $\ep\ll 1$  
at the beginning, so it can serve as an
expansion parameter to construct an analytic solution of 
the system (\ref{2})-(\ref{7}) to describe the dependence of
 all the above variables 
on the initial moment at which we set initial conditions. For 
the sake of brevity,  below we will call this dependence
``time evolution''.

For our analytic expansion, it is convenient to rewrite the system of 
equations (\ref{2})-(\ref{7})
 in terms of the tilde-variables, all of which tend to $1$ at spatial infinity.
From (\ref{4}), we find
\be
\frac{\tilde{a}'}{\tilde{a}}+\frac{\gamma}{1+\gamma}\frac{\tilde{\rho}'}{\tilde{\rho}}
=[\ln(\tilde{a}\tilde{\rho}^{\frac{\gamma}{1+\gamma}})]'=0,
\ee
so
\be
\tilde{a}=F(t)\tilde{\rho}^{-\frac{\gamma}{1+\gamma}},
\ee
where $F(t)$ is an arbitrary function of time. For convenience we choose 
$F(t)=1$, then
\be
\tilde{a}=\tilde{\rho}^{-\frac{\gamma}{1+\gamma}}.\label{23}
\ee
Such a choice of $F(t)$ corresponds to a frame of reference which is 
synchronous at spatial infinity.

Since both $\ti{\rho}$ and $\ti{a}$ are positive definite, we may define $\hat{\rho}\equiv \ln{\ti{\rho}}$
and $\hat{a}\equiv \ln{\ti{a}}$ and then rewrite (\ref{23}) as
\be
\hat{a}=-\fr{\ga}{1+\ga}\hat{\rho}.\label{hata}
\ee
Using the tilde-variables, we can rewrite (\ref{2}) as
\be
\al\tilde{R}+t\dot{\tilde{R}}=\al \tilde{a}\tilde{H}\tilde{R},
\ee
so
\be
t\dot{\tilde{R}}=\al\tilde{R}(\tilde{\Phi}-1),\label{26}
\ee
where
\be
\tilde{\Phi}\equiv \tilde{a}\tilde{H}.
\ee
Since $\tilde{R}$ is positive, we can also define $\hat{R}$ by $\hat{R}\equiv\ln\ti{R}$.
Introducing a new time variable
\be
\xi \equiv \ln\left(\fr{t}{t_\mathrm{i}}\right),
\ee
(\ref{26}) is expressed as
\be
\frac{\partial \hat{R}}{\partial \xi}=\al(\tilde{\Phi}-1).\label{30}
\ee

From the definition of the curvature profile function $K(r,t)$, (\ref{16}), we find
\be
-r^2\dot{K}=\left(\frac{R'^2}{b^2}\right)^\cdot=\frac{2R'^2}{b^2}\left(\frac{\dot{R}'}{R'}-\frac{\dot{b}}{b}\right)
=2(1-Kr^2)\left(\frac{\dot{R}'}{R'}-\frac{\dot{b}}{b}\right).
\ee
Using (\ref{2}) and (\ref{3}), this can be rewritten as
\be
-r^2\dot{K}
=2H(1-Kr^2)\frac{a'r\tilde{R}}{(r\tilde{R})'}\equiv 2H(1-Kr^2)D_r\tilde{a},\label{31}
\ee
where we have introduced the operator
\be
D_r\equiv \frac{r\tilde{R}\pa}{(r\tilde{R})'\pa r}.
\ee
We write the initial condition for (\ref{31}) as
\be
K(0,r)\equiv K_\mathrm{i}(r),
\ee
where $K_\mathrm{i}(r)$ is an arbitrary function of $r$ which vanishes outside the perturbed region,
and define a new variable $\tilde{K}$ by
\be
1-K(t,r)r^2=(1-K_\mathrm{i}(r)r^2)\tilde{K}(t,r).\label{34}
\ee
$\tilde{K}$ is unity at spatial infinity like the other tilde-variables, but
in contrast to the other tilde-variables,
it describes the evolution of curvature deviation from the initial curvature
profile rather than the deviation from the spatially flat Friedmann universe. Note that,
from the definition (\ref{16}) of $b(t,r)$, $K_\mathrm{i}(r)$ has to satisfy the condition
\be
K_\mathrm{i}(r)<\fr{1}{r^2}.
\ee
Physically, this condition ensures that the perturbed region does not form a closed universe 
which is causally disconnected from our universe \cite{Carr:2010har,PhysRevD.83.124025}.

Differentiating (\ref{34}) with respect to $t$ and using (\ref{31}), we find
\be
(1-K_\mathrm{i}r^2)\dot{\tilde{K}}=-r^2\dot{K}=2H(1-Kr^2)D_r\tilde{a}=2H(1-K_\mathrm{i}r^2)\tilde{K}D_r\tilde{a},
\ee
which yields
\be
t\dot{\tilde{K}}=2\al\tilde{H}\tilde{K}D_r\tilde{a}.\label{36}
\ee
Since $\tilde{K}$ is always positive, we can define another hat variable, $\hat{K}\equiv\ln \ti{K}$,
and rewrite (\ref{36}) using (\ref{23}) as
\be
\frac{\pa \hat{K}}{\pa \xi}=2\al \tilde{H}D_r\tilde{a}
=-\frac{2\al \gamma \tilde{\Phi}}{1+\gamma}\frac{D_r\tilde{\rho}}{\tilde{\rho}}
=-\frac{4\gamma}{3(1+\gamma)^2}\tilde{\Phi}D_r\hat{\rho}.\label{38}
\ee
%hence
%\be
%\frac{\pa \hat{K}}{\pa\xi}=
%\ee
Using (\ref{17}), we can write (\ref{5}) in terms of the tilde-variables as
\be
\frac{4\pi}{3}\rho_0R^3\tilde{\mu}\left(\frac{\tilde{\mu}'}{\tilde{\mu}}
+3\frac{(r\tilde{R})'}{r\tilde{R}}\right)=4\pi \rho R^3\frac{(r\tilde{R})'}{r\tilde{R}},
\ee
which leads to
\be
\tilde{\mu}'+3\tilde{\mu}\frac{(r\tilde{R})'}{r\tilde{R}}=3\tilde{\rho}\frac{(r\tilde{R})'}{r\tilde{R}}
\ee
and hence
\be
\tilde{\rho}=\tilde{\mu}+\frac{1}{3}D_r\tilde{\mu}.\label{39}
\ee

In terms of the tilde-variables, (\ref{6}) can be written as
\be
\dot{\tilde{\mu}}+\tilde{\mu}\left(-\frac{2}{t}+3\frac{\dot{R}}{R}\right)
+3\gamma\frac{\rho}{\rho_0}\frac{\dot{R}}{R}=0,
\ee
so
\be
t\dot{\tilde{\mu}}=2\tilde{\mu}-3\al \tilde{a}\tilde{H}(\tilde{\mu}+\gamma\tilde{\rho}).\label{43}
\ee
Defining
\be
\tilde{f}\equiv \frac{\tilde{\mu}+\gamma \tilde{\rho}}{1+\gamma},
\ee
(\ref{43}) can be rewritten as
\be
\frac{\pa \tilde{\mu}}{\pa \xi}=2(\tilde{\mu}-\tilde{\Phi}\tilde{f}).\label{44}
\ee
In terms of $H$, the constraint equation (\ref{7}) is expressed as
\be
H^2=\frac{8\pi G}{3}\rho_0\tilde{\mu}-\frac{Kr^2}{R^2}\label{46}
\ee
and this gives
\be
\tilde{\mu}=\tilde{H}^2+\frac{\ep Kr_\mathrm{i}^2}{\tilde{R}^2}.\label{47}
\ee

One important property of the perturbation follows from (\ref{46}) and the boundary conditions. 
The equation (\ref{46}) corresponds to the Friedmann equation of the flat FLRW universe
\be
H_0^2=\fr{8\pi G}{3}\rho_0.\label{Friedmann}
\ee
Using (\ref{defofM}) and (\ref{17}), (\ref{46}) and (\ref{Friedmann}) are combined to give
\be
\fr{H^2-H_0^2}{H_0^2}=4\pi\int^R_0\left(\fr{\rho-\rho_0}{\rho_0}\right)R^2dR-\fr{Kr^2}{R^2H_0^2}.\label{difference}
\ee
Noting the left-hand side and the second term of the right-hand side vanish at spatial infinity 
as a result of the boundary conditions and defining the energy density perturbation
\be
\de(t,r)\equiv \fr{\rho(t,r)-\rho_0(t)}{\rho_0(t)}=\ti{\rho}(t,r)-1,\label{defofdelta}
\ee
(\ref{difference}) leads to the following condition for $\de$:
\be
4\pi\int^\infty_0\de R^2dR=0.\label{condfordelta}
\ee
Namely, the mass excess in the center has to be compensated by the sorrounding mass 
deficit in order for the solution to coinside with the flat FLRW solution at spatial infinity.

Equations (\ref{23}), (\ref{30}), (\ref{38}), (\ref{39}), (\ref{44}) and (\ref{47}) are the fundamental equations to solve.
%%%%%%%%%%%%%%%%%%%%%%%%%%%%%%%%%%%%%%%%%%%%%%%%%%%%%%%
\section{Expansion over $\ep$ and quasi-linearization}
%%%%%%%%%%%%%%%%%%%%%%%%%%%%%%%%%%%%%%%%%%%%%%%%%%%%%%%
We now expand the tilde-variables over the parameter $\ep$ as a first step to solving these 
fundamental equations:
\be
\tilde{X}(t,r)=\sum ^\infty_{n=0}\ep^n(t)\tilde{X}_{(n)}(r).
\ee
Note that by definition $\tilde{X}_{(0)}=1$ and $\ti{X}'_{(0)}=0$.
The hat-variables are expanded similarly. 
Differentiating this expansion with respect to $t$, one obtains
\be
\dot{\tilde{X}}(t,r)=\dot{\ep}\sum^\infty_{n=0}n\ep^{n-1}(t)\tilde{X}_{(n)}(r)
=\frac{\beta}{t}\sum^\infty_{n=1}n\ep^n(t)\tilde{X}_{(n)}(r),
\ee
hence
\be
\left(\dot{\tilde{X}}\right)_{(n)}=\frac{n\beta}{t}\tilde{X}_{(n)}.\label{54}
\ee
Differentiating the expansion with respect to $r$, one finds
\be
\tilde{X}'(t,r)=\sum^{\infty}_{n=1}\ep^n(t)\tilde{X}_{(n)}'(r),
\ee
hence
\be
(\tilde{X}')_{(n)}=\tilde{X}'_{(n)}.\label{56}
\ee

Let us consider the product of two tilde-variables $\ti{X}_1$ and $\ti{X}_2$:
\be
\tilde{X}_1\tilde{X}_2=\left(\sum^\infty_{i=0}\ep^i\tilde{X}_{1(i)}\right)
\left(\sum^\infty_{j=0}\ep^j\tilde{X}_{2(j)}\right)
=\sum^\infty_{i=0}\sum^\infty_{j=0}\ep^{i+j}\tilde{X}_{1(i)}\tilde{X}_{2(j)},
\ee
from which one finds
\be
(\tilde{X}_1\tilde{X}_2)_{(n)}=\sum^n_{i=0}\tilde{X}_{1(i)}\tilde{X}_{2(n-i)}.
\ee
Since $\tilde{X}_{1(0)}=\tilde{X}_{2(0)}=1$, one obtains
\be
(\tilde{X}_1\tilde{X}_2)_{(n)}=\tilde{X}_{1(n)}+\tilde{X}_{2(n)}
+S_{(n)}[\tilde{X}_1\tilde{X}_2],\label{57}
\ee
where
\be
S_{(n)}[\tilde{X}_1\tilde{X}_2]\equiv \sum^{n-1}_{i=1}\tilde{X}_{1(i)}\tilde{X}_{2(n-i)}.
\ee
Note that $S_{(0)}[\tilde{X}_1\tilde{X}_2]=S_{(1)}[\tilde{X}_1\tilde{X}_2]=0$.
The most important feature of $S_{(n)}[\ti{X_1}\ti{X_2}]$ is that it depends only on 
coefficients up to $(n-1)$th order.

The relationship (\ref{57}) can be generalized to arbitrary functions of the tilde-variables. 
Let $F_1$ and $F_2$ be arbitrary functions of tilde-variables
and their time and space derivatives. Then 
\be
(F_1F_2)_{(n)}=\sum^n_{i=0}F_{1(i)}F_{2(n-i)}=F_{1(n)}F_{2(0)}+F_{1(0)}F_{2(n)}+S_{(n)}[F_1F_2].\label{59}
\ee
As a consequence of (\ref{56}) and (\ref{59}), since $\tilde{X}'_{i(0)}=0$, one finds
\be
(\tilde{X}'F)_{(n)}=\tilde{X}'_{(n)}F_{(0)}+S_{(n)}[(\tilde{X})'F],\label{61}
\ee
and
%where $F$ is an arbitrary function of the tilde-variables and their time and space derivatives. We also obtain
\be
(\tilde{X_1}'\tilde{X_2}')_{(n)}=S_{(n)}[(\tilde{X_1}')(\tilde{X_2}')].
\ee
Similarly, using (\ref{54}) and (\ref{59}) and noting $\dot{\tilde{X}}_{(0)}=0$, one obtains
\be
(\dot{\tilde{X}}F)_{(n)}=\frac{n\beta}{t}\tilde{X}_{(n)}F_{(0)}+S_{(n)}[(\dot{\tilde{X}})F]
=\frac{n\beta}{t}\left(\tilde{X}_{(n)}F_{(0)}+\frac{1}{n}S^*_{(n)}[\tilde{X}F]\right),
\ee
where we have defined
\be
S^*_{(n)}[\tilde{X}F]=\sum^{n-1}_{m=1}m\tilde{X}_mF_{(n-m)}.
\ee
We also find
\be
(\dot{\tilde{X_1}}\tilde{X_2}')_{(n)}=\frac{\beta}{t}S^*_{(n)}[\tilde{X_1}\tilde{X_2}'].
\ee

It is useful to obtain relationships between the expansion coefficients of the tilde-variables
and those of the hat-variables. Suppose $\ti{Y}$ is some product of the positive-definite quantities $\ti{a},\ti{R},\ti{\rho}$ and $\ti{K}$, such as
\be
\ti{Y}\equiv\ti{a}^{p_1}\ti{R}^{p_2}\ti{\rho}^{p_3}\ti{K}^{p_4},
\ee
where $p_1,\cdots, p_2$ are integers. Then defining
\be
\hat{Y}\equiv \ln\ti{Y}=p_1\hat{a}+p_2\hat{R}+p_3\hat{\rho}+p_4\hat{K},
\ee
we can relate the expansion coefficients of %$\ti{Y}$ and $\hat{Y}$, which are formally defined by
\be
\ti{Y}_{(n)}=\fr{1}{n!}\lim_{\ep \to 0}\fr{\pa^n\ti{Y}}{\pa\ep^n}
\ee
and
\be
\hat{Y}_{(n)}=\fr{1}{n!}\lim_{\ep \to 0}\fr{\pa^n\hat{Y}}{\pa\ep^n},
\ee
by
\begin{eqnarray}
\ti{Y}_{(n)}%&=&\fr{1}{n!}\lim_{\ep \to 0}\fr{\pa^n}{\pa \ep^n}e^{\hat{Y}} \nonumber \\
&=&\fr{1}{n}\sum^n_{m=1}\fr{m!}{(m-1)!}\lim_{\ep \to 0}\left(\fr{\pa^m\hat{Y}}{m!\pa \ep^m}\right)
\lim_{\ep \to 0}\left(\fr{\pa^{n-m}}{(n-m)!\pa \ep^{n-m}}e^{\hat{Y}}\right) \nonumber \\
&=&\fr{1}{n}\sum^n_{m=1}m\hat{Y}_{(m)}\ti{Y}_{(n-m)}.
\end{eqnarray}
Using $\ti{Y}_{(0)}=1$, we obtain
\be
\ti{Y}_{(n)}=\hat{Y}_{(n)}+\fr{1}{n}S^*_{(n)}[\hat{Y}\ti{Y}].\label{73}
\ee

\section{Equations for analytic calculations}
The fundamental equations to be solved in the following are 
(\ref{hata}), (\ref{30}), (\ref{38}), (\ref{39}), (\ref{44}) and (\ref{47}).
%In section 2, the following equations for the tilde and hat variables 
%were obtained.
%\be
%\tilde{\mu}=\tilde{H}^2+\ep\left[\fr{r_\mathrm{i}^2}{r^2}+e^{\hat{K}}\left(K_\mathrm{i}r_\mathrm{i}^2-\fr{r_\mathrm{i}^2}{r^2}\right)\right]e^{-2\hat{R}},\label{74}
%\ee
%\be
%\frac{\pa \tilde{\mu}}{\pa \xi}=2(\tilde{\mu}-\tilde{\Phi}\tilde{f}),\label{75}
%\ee
%\be
%\tilde{\rho}=\tilde{\mu}+\frac{1}{3}D_r\tilde{\mu},\label{76}
%\ee
%\be
%\hat{a}=-\frac{\gamma}{1+\gamma}\hat{\rho},\label{77}
%\ee
%\be
%\frac{\pa \hat{R}}{\pa \xi}=\fr{2}{3(1+\ga)}(\tilde{\Phi}-1),\label{78}
%\ee
%\be
%\frac{\pa\hat{K}}{\pa \xi}=-\frac{4\gamma}{3(1+\gamma)^2}\tilde{\Phi}D_r\hat{\rho}.\label{79}
%\ee
We solve for the expansion coefficients $X_{i(n)}$ of each tilde or hat variable in the power series 
expansion with respect to $\ep$ using these equations. To do this, we derive a set of recursive formulae 
to express $X_{i(n)}$ in terms of $X_{j(m)}$ with $m<n$.

First from (\ref{hata}) we find 
\be
\hat{a}_{(n)}=-\fr{\ga}{1+\ga}\hat{\rho}_{(n)},
\ee 
which yields
\be
(1+\ga)\ti{a}_{(n)}+\ga\ti{\rho}_{(n)}=\fr{\ga}{n}S_{(n)}^*[\hat{\rho}(\ti{\rho}-\hat{a})].\label{i}
\ee
From (\ref{44}) and (\ref{47}) with (\ref{i}), we can express $\ti{H}_{(n)}$ and $\ti{\mu}_{(n)}$ in terms of the lower-order coefficients as follows.
Using (\ref{57}), (\ref{59}) and (\ref{73}), (\ref{47}) leads to
\begin{eqnarray}
\ti{\mu}_{(n)}&-&2\ti{H}_{(n)}
=S_{(n)}[\ti{H}\ti{H}]+\fr{r_\mathrm{i}^2}{r^2}(e^{-2\hat{R}})_{(n-1)}\nonumber\\
&+&r_\mathrm{i}^2\left(K_\mathrm{i}-\fr{1}{r^2}\right)\left\{(e^{\hat{K}})_{(n-1)}+(e^{-2\hat{R}})_{(n-1)}+S_{(n-1)}[e^{\hat{K}}e^{-2\hat{R}}]\right\}\nonumber\\
&\equiv& F_{(n)}+W_{1(n)}\label{ro}
\end{eqnarray}
where
\be
F_{(n)}\equiv\delta^1_nr_i^2K_i-2r_\mathrm{i}^2K_\mathrm{i}\hat{R}_{(n-1)}+r_\mathrm{i}^2\left(K_\mathrm{i}-\fr{1}{r^2}\right)\hat{K}_{(n-1)},
\ee
and
\begin{eqnarray}
W_{1(n)}&\equiv&S_{(n)}[\ti{H}\ti{H}]+r_\mathrm{i}^2\left(K_\mathrm{i}-\fr{1}{r^2}\right)S_{(n-1)}[e^{\hat{K}}e^{-2\hat{R}}]\nonumber\\
&+&\fr{1}{n-1}\left\{r_\mathrm{i}^2\left(K_\mathrm{i}-\fr{1}{r^2}\right)S^*_{(n-1)}[\hat{K}e^{\hat{K}}]-2r_\mathrm{i}^2K_\mathrm{i}S_{(n-1)}^*[\hat{R}e^{-2\hat{R}}]\right\}.
\end{eqnarray}
On the other hand, (\ref{44}) yields
\be
n\beta\ti{\mu}_{(n)}=2\ti{\mu}_{(n)}-2(\ti{\Phi}\ti{f})_{(n)}.
\ee
Using the equalities
\be
\ti{\Phi}_{(n)}=\ti{a}_{(n)}+\ti{H}_{(n)}+S_{(n)}[\ti{a}\ti{H}],
\ee
\be
\ti{f}_{(n)}=\fr{1}{1+\ga}(\ga\ti{\rho}_{(n)}+\ti{\mu}_{(n)}),
\ee
we find
\begin{eqnarray}
(\ti{\Phi}\ti{f})_{(n)}
&=&\ti{a}_{(n)}+\ti{H}_{(n)}+S_{(n)}[\ti{a}\ti{H}]+\fr{1}{1+\ga}(\ti{\mu}_{(n)}+\ga\ti{\rho}_{(n)})+S_{(n)}[\ti{\Phi}\ti{f}]\nonumber\\
&=&\ti{H}_{(n)}+\fr{1}{1+\ga}\ti{\mu}_{(n)}+\fr{\ga}{(1+\ga)n}S_{(n)}^*[\hat{\rho}(\ti{\rho}-\ti{a})]\nonumber\\
&+&S_{(n)}[\ti{a}\ti{H}]+S_{(n)}[\ti{\Phi}\ti{f}],\label{ha}
\end{eqnarray}
where we have used (\ref{i}) in the last equality.
From (\ref{ro}) and (\ref{ha}) we have%can express $\ti{\mu}_{(n)}$ and $\ti{H}_{(n)}$ as
\be
\ti{\mu}_{(n)}=\fr{1}{1+A_n}(F_{(n)}+W_{1(n)}-W_{2(n)}),\label{mutilden}
\ee
\be
\ti{H}_{(n)}=-\fr{1}{2(1+A_n)}[A_n(F_{(n)}+W_{1(n)})+W_{2(n)}],\label{Htilden}
\ee
where
\be
A_n\equiv \fr{2}{1+\ga}\left[\left(\ga+\fr{1}{3}\right)n-\ga\right],
\ee
\be
W_{2(n)}\equiv 2\left(S_{(n)}[\ti{a}\ti{H}]+S_{(n)}[\ti{\Phi}\ti{f}]+\fr{\ga}{n(1+\ga)}S_{(n)}^*[\hat{\rho}(\ti{\rho}-\ti{a})]\right).
\ee

Now that we have expressed $\ti{\mu}_{(n)}$ and $\ti{H}_{(n)}$ in terms of lower-order coefficients, we may use these coefficients
to obtain recursive formulae for the other variables.
For example, from (\ref{39}) we find% a formula for $\ti{\rho}_{(n)}$ as
\be
\ti{\rho}_{(n)}=\ti{\mu}_{(n)}+\fr{r}{3}\ti{\mu}_{(n)}'+W_{3(n)},\label{rhotilden}
\ee
with
\be
W_{3(n)}\equiv S_{(n)}[(\ti{\mu}-\ti{\rho})(r\ti{R})']+\frac{r}{3}S_{(n)}[\ti{\mu}'\ti{R}],
\ee
where we have used (\ref{61}). Then from (\ref{i}) we find
\be
\ti{a}_{(n)}=-\fr{\ga}{1+\ga}\left(\ti{\mu}_{(n)}+\fr{r}{3}\ti{\mu}'_{(n)}\right)+W_{4(n)},\label{atilden}
\ee
with
\be
W_{4(n)}\equiv -\fr{\ga}{1+\ga}W_{3(n)}+\fr{\ga}{(1+\ga)n}S_{(n)}^*[\hat{\rho}(\ti{\rho}-\hat{a})].
\ee
Similarly (\ref{30}) yields
\be
\hat{R}_{(n)}=\fr{1}{(1+3\ga)n}(\ti{a}_{(n)}+\ti{H}_{(n)}+W_{5(n)}),
\ee
with
\be
W_{5(n)}\equiv S_{(n)}[\ti{a}\ti{H}].
\ee
From (\ref{38}) 
\be
\hat{K}_{(n)}=-\fr{2\ga}{(1+\ga)(1+3\ga)n}(r\hat{\rho}_{(n)}'+W_{6(n)}),
\ee
with
\be
W_{6(n)}\equiv rS_{(n)}[\hat{\rho}'(\ti{\Phi}\ti{R})]+\left(2+\fr{1}{2\ga}+\fr{3\ga}{2}\right)S_{(n)}^*[\hat{K}(r\ti{R})'].
\ee
The corresponding tilde-variables are obtained from

\be
\ti{R}_{(n)}=\hat{R}_{(n)}+\fr{1}{n}S_{(n)}^*[\hat{R}\ti{R}],\label{Rtilden}
\ee
\be
\ti{K}_{(n)}=\hat{K}_{(n)}+\fr{1}{n}S_{(n)}^*[\hat{K}\ti{K}].\label{Ktilden}
\ee
This completes our derivation of the recursive formulae.
The first-order coefficients are determined by the initial profile of the curvature inhomogeneity $K_\mathrm{i}(r)$ as

\be
\ti{\mu}_{(1)}(r)=\fr{3(1+\ga)r_\mathrm{i}^2K_\mathrm{i}(r)}{5+3\ga},
\ee
\be
\ti{H}_{(1)}(r)=-\fr{r_\mathrm{i}^2K_\mathrm{i}(r)}{5+3\ga},
\ee
\be
\ti{\rho}_{(1)}(r)=\fr{(1+\ga)r_\mathrm{i}^2(3K_\mathrm{i}(r)+rK_\mathrm{i}'(r))}{5+3\ga}, \label{rho1}
\ee
\be
\ti{a}_{(1)}(r)=-\fr{\ga r_\mathrm{i}^2(3K_\mathrm{i}(r)+rK_\mathrm{i}'(r))}{5+3\ga},
\ee
\be
\ti{R}_{(1)}(r)=-\fr{(1+3\ga)r_\mathrm{i}^2K_\mathrm{i}(r)+\ga r_\mathrm{i}^2 r K_\mathrm{i}'(r)}{5+18\ga +9\ga^2},
\ee
\be
\ti{K}_{(1)}(r)=-\fr{2\ga r_\mathrm{i}^2r(4K_\mathrm{i}'(r)+rK_\mathrm{i}''(r))}{(1+3\ga)(5+3\ga)}.
\ee

%\section{Preliminary analysis of required order in asymptotic expansion}

\section{Analytic solution}
Dropping all terms of order 
greater than $N$, we obtain truncated asymptotic solutions of $N$th order.  
In order to use these solutions to set the initial conditions of numerical simulation of PBH formation, 
it is necessary to estimate an upper limit on time when such solutions are accurate enough to be used. 
Let the maximum acceptable error of the analytic solution be $\Delta$. 
Then the latest epoch for which the analytic solution is
accurate enough 
is determined by 
the first dropped terms of the truncated asymptotic expansions. 
If we use the asymptotic expansion of $N$th order, the error of the asymptotic expansion for $\tilde{X}$ is\\ 
\be
\mathrm{ERR}\left(\sum^N_{n=0}\ep^n\tilde{X}_{(n)}\right)
\equiv\tilde{X}-\sum^N_{n=0}\ep^n\tilde{X}_{(n)}=O(\ep^{N+1}\tilde{X}_{(N+1)}).
\ee
Then we require
\be
\ep^{N+1}M_{(N+1)}<\Delta,
\ee
where $M_{(n)}$ and $M_{\ti{X}_{(n)}}$ are defined by
\be
M_{(n)}\equiv \max\left\{
M_{\ti{a}_{(n)}},M_{\ti{R}_{(n)}},M_{\ti{K}_{(n)}},
M_{\ti{\rho}_{(n)}},M_{\ti{\mu}_{(n)}},M_{\ti{H}_{(n)}}
\right\}
\ee
and
\be
M_{\ti{X}_{(n)}}\equiv\mathrm{\max_\mathit{r}}|\ti{X}_{(n)}(r)|,
\ee
respectively.
The error associated with the
analytic calculation is less than $\Delta$ if it is calculated when
\be
\ep<\ep_{\mathrm{max}}\equiv \sqrt[N+1]{\fr{\Delta}{M_{(N+1)}}}.
\ee

We solve the recursive relations obtained in \S I\hspace{-.1em}V for 
four specific curvature profiles of the form
\be
K_\mathrm{i}(r)=\left[1+\fr{B}{2} \left(\frac{r}{\sigma}\right)^2\right]
\exp\left[-\fr{1}{2}\left(\frac{r}{\sigma}\right)^2\right], \label{Kini}
\ee
where $B$ describes slope of
curvature profiles and $\sigma$ specifies the comoving
length scale of curvature profile.
Smaller values of $B$ correspond to shallower profiles,
and when $B=0$ the profile is simply Gaussian. 
The amplitude of the profile is set to unity at the origin
where the same normalization is used as a spatially closed Friedmann universe
in accordance with \cite{Polnarev:2006aa}.

In order to represent the comoving length scale of the 
perturbed region,
we use the comoving radius,
$r_\mathrm{i}$, of the overdense region.
We can calculate $r_{\mathrm i}$ by %defining energy density perturbation as
%\be
%\delta(t,r)\equiv\fr{\rho(t,r)-\rho_0(t)}{\rho_0(t)}=\ti{\rho}(t,r)-1
%\ee
%and 
solving the following equation for the energy density perturbation defined by (\ref{defofdelta}):
\be
\delta(t,r_{\mathrm i})=0.
\ee
Since the initial condition is taken at the superhorizon regime, when
$\epsilon$ is extremely small, the lowest-order solution
(\ref{rho1}) suffices to calculate $r_\mathrm{i}$, which is obtained by solving 
\be
3K_\mathrm{i}(r_\mathrm{i})+r_{\mathrm i}K'_\mathrm{i}(r_\mathrm{i})=0. 
\label{rieq}
\ee
With the current choice of the functional form of $K_\mathrm{i}(r)$, 
(\ref{Kini}),
the solution of (\ref{rieq})  is given by
\be
r_\mathrm{i}^2=3\sigma^2~~{\rm for}~~B=0,
\ee
and
\be
r_\mathrm{i}^2=\sigma^2\frac{5B-2+\sqrt{(5B-2)^2+24B}}{2B}~~{\rm
for}~~B\neq 0.
\ee

\begin{figure}[h]

\begin{center}
\includegraphics[width=14cm,keepaspectratio,clip]{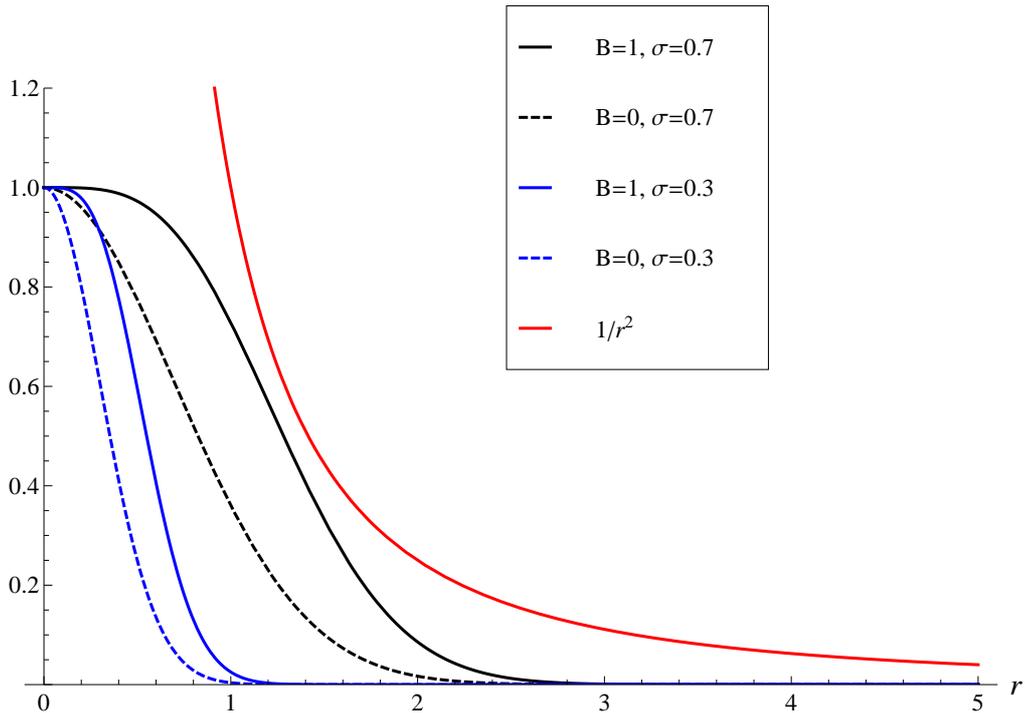}
\end{center}
\caption{Initial curvature profiles $K_\mathrm{i}(r)$ which are used as specific examples to obtain 
expansion coefficients of the tilde-variables. 
%The parameters of these profiles are $(B,\sigma)=(1,0.7),(0,0.7),(1,0.3),(0,0.3)$.
Note that these functions have to satisfy $K_\mathrm{i}(r)<1/r^2$.}
\label{profiles}
\end{figure}

%We normalize radial coordinate so that $r_\mathrm{i}$ takes unity in
%the 
%following computation.
We have obtained analytic solutions for curvature profiles with 
$(B,\sigma)=(1,0.7),(0,0.7),(1,0.3),(0,0.3)$, corresponding to
wide and steep, wide and shallow, narrow and steep, narrow and shallow
profiles, 
respectively. Plots of these profiles are shown in
Figure \ref{profiles}.  Note that the physical length scale in the
asymptotic Friedmann region
is obtained by multiplying by the scale factor $S(t)$, whose
normalization
we have not specified.  We can therefore  set up initial 
conditions for PBH formation with arbitrary mass scales by adjusting
the normalization of $S(t)$ which appears in the expansion parameter.

\begin{figure}[h]
\begin{center}
\includegraphics[width=14cm,keepaspectratio,clip]{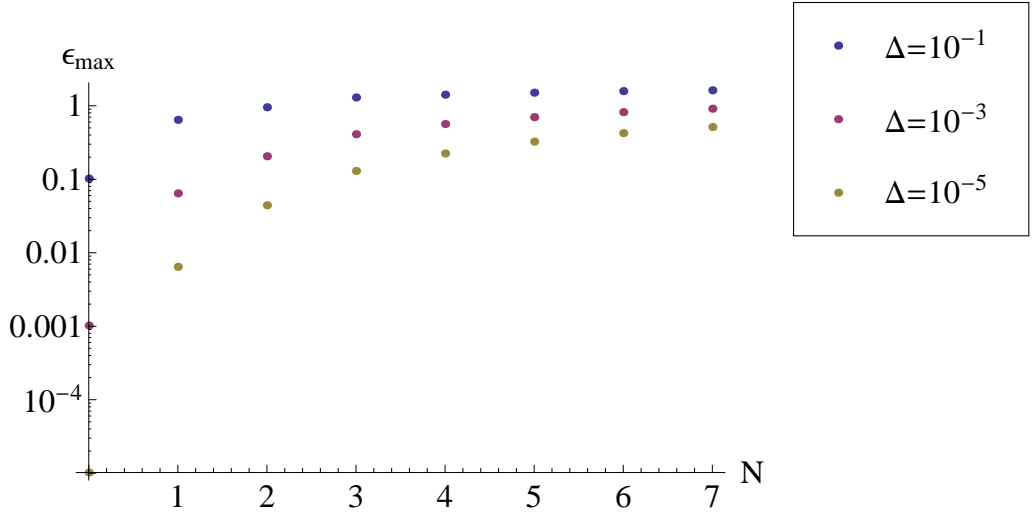}
\end{center}
\caption{The dependence of $\ep_\mathrm{max}$ on the order of the asymptotic expansion, $N$, 
is shown. The profile with $(B,\sigma)=(0,0.7)$ 
is used.% and $\Delta$ is set to be $10^{-1}, 10^{-3}$ and $10^{-5}$.
}
\label{epmax_order_007QM}
\end{figure}

\begin{figure}[h]
\begin{center}
\includegraphics[width=14cm,keepaspectratio,clip]{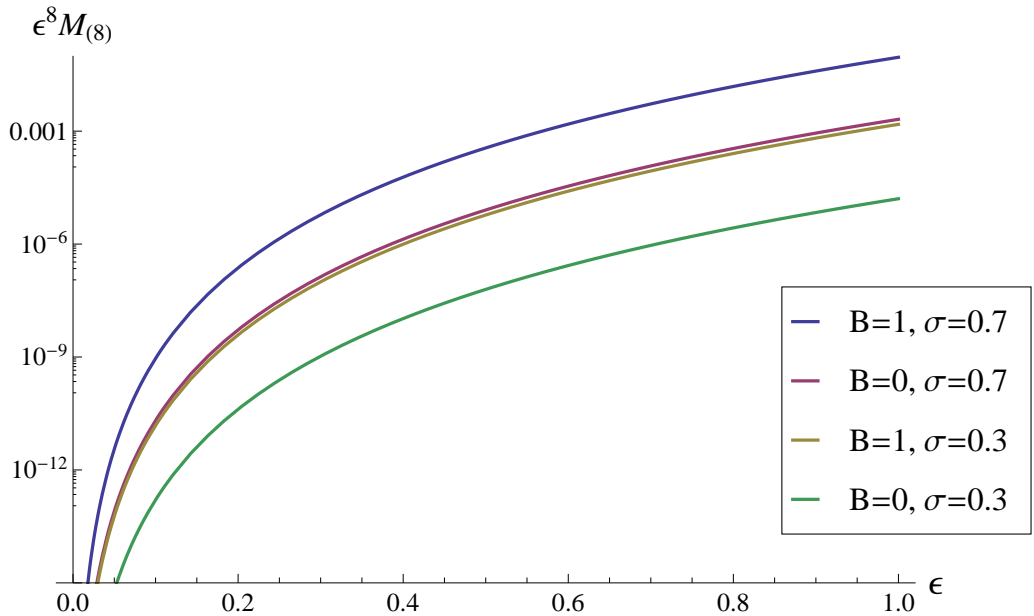}
\end{center}
\caption{The time dependence of errors associated with analytic solution obtained by asymptotic expansion of seventh order.
%When initial curvature fluctuation is wider and its profile is steeper, expansion coefficients tend to become larger,
%hence errors of analytic solution are relatively larger.
}
\label{errors-8th-order-new}
\end{figure}

For these four specific profiles, expansion coefficients of the 
tilde-variables are calculated by solving the recurrence formulae 
(\ref{mutilden}), (\ref{Htilden}),
(\ref{rhotilden}), (\ref{atilden}), (\ref{Rtilden}) and (\ref{Ktilden}) 
numerically. 
Then, the quantity $\ep_{\mathrm{max}}$  
was calculated for $\Delta=10^{-1},10^{-3},10^{-5}$ and $N=1-7$. 
The values of $\ep_{\mathrm{max}}$ are summarized in Table \ref{table}.
When an asymptotic expansion of higher-order is used, $\ep_{\mathrm{max}}$ 
is larger, so the analytic solution constructed is
sufficiently accurate until a later time.
For instance, one can see from the table that 
when an asymptotic expansion of first order is used, 
the numerical calculation 
has to be started at $\ep=0.0064$ in order to maintain the accuracy 
of order $10^{-5}$, for the profile with $(B,\sigma)=(0,0.7)$. 
On the other hand, 
if we use an asymptotic expansion of seventh order, 
we can follow the evolution of perturbation until 
$\ep=0.51$, maintaining the accuracy of $10^{-5}$, 
for the profile with $(B,\sigma)=(0,0.7)$. 
The dependence of $\ep_\mathrm{max}$ on the order of the asymptotic expansion, $N$, 
is more clearly seen from Figure \ref{epmax_order_007QM}, in which the profile with $(B,\sigma)=(0,0.7)$ 
is used and $\Delta$ is set to be $10^{-1}, 10^{-3}$ and $10^{-5}$.
The time dependence of $\Delta$ for asymptotic expansion of
seventh order is shown in Figure 
\ref{errors-8th-order-new}.
Note that it is determined by the first dropped eighth order terms of the expansions in this case.
When the initial curvature fluctuation is wider and its profile steeper, expansion coefficients tend to 
be larger,
so the errors in the analytic solution are also larger.

\begin{figure}[h]
\begin{center}
\includegraphics[width=14cm,keepaspectratio,clip]{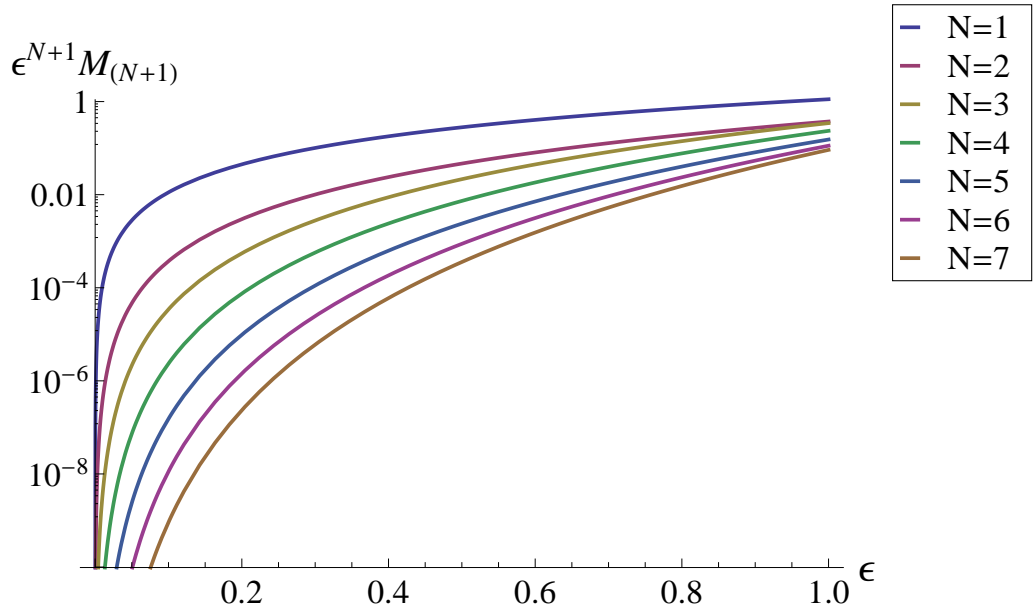}
\end{center}
\caption{Comparison of the time dependence of errors associated with analytic solutions obtained by 
asymptotic expansions of different orders. 
The profile with $(B,\sigma)=(1,0.7)$ was used for these plots. 
%One can see clearly that errors with higher-order expansions are relatively small and increase slowly 
%in comparison to those with lower-order expansions.
}
\label{errors_different_orders}
\end{figure}

\begin{figure}[h]
\begin{center}
\includegraphics[width=17cm,keepaspectratio,clip]{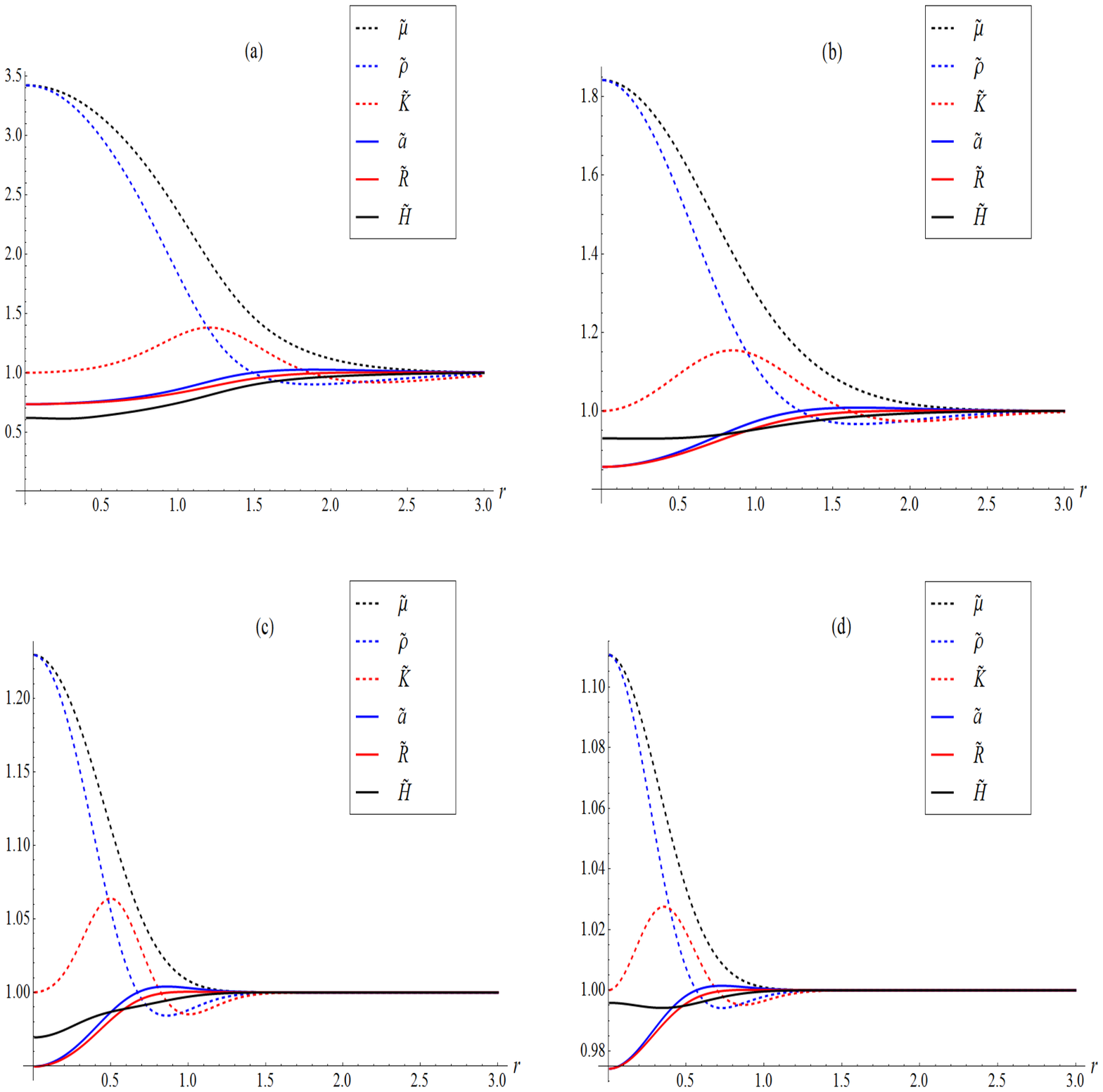}
\end{center}
\caption{Profiles of the tilde-variables at $\ep$=0.9, which were calculated from an asymptotic expansion of seventh order.
(a), (b), (c) and (d) were obtained from $(B,\sigma)=(1,0.7),(0,0.7),(1,0.3)$, and $(0,0.3)$, respectively.
Note that the errors associated with these profiles are less than of order $10^{-3}$ from Figure \ref{errors-8th-order-new}.}
\label{tildevariables}
\end{figure}

Comparison of the time dependence of the errors associated with analytic
solutions with different orders 
is shown in Figure \ref{errors_different_orders}. 
The profile with $(B,\sigma)=(1,0.7)$ was used for these plots. 
One can see clearly that the errors with higher-order expansions 
are relatively small and increase more slowly 
than those with lower-order expansions. 
Plots of the tilde-variables at $\ep=0.9$, calculated using the asymptotic expansion of seventh order, is shown in Figure \ref{tildevariables}.
Note that errors associated with these plots are less than $10^{-3}$ from Figure \ref{errors-8th-order-new}.
\begin{table}
\begin{tabular}{|c|c|c|c|c|c|c|c|c|c|} \hline
 $(B,\sigma)$&\backslashbox{$\Delta$}{$N$}  &$0$ &$1$ &$2$ &$3$ &$4$ &$5$ &$6$ &$7$ \\ \hline
 &$10^{-1}$&0.070      &   0.30   &    0.64  &   0.73   &   0.84   &  0.93    &    0.98  & 1.0 \\ \cline{2-10}
 $(1,0.7)$&$10^{-3}$&$7.0\times 10^{-4} $     &0.030     &0.14      & 0.23     &0.34      &0.43      &0.51      &0.57 \\ \cline{2-10}
 &$10^{-5}$&$7.0\times 10^{-6}$      &0.0030      &0.030      &0.073      &0.13      &0.20      &0.26      &0.32 \\ \hline
 &$10^{-1}$&0.10      &   0.64   &    0.95  &   1.3   &   1.4   &  1.5    &    1.6  & 1.6 \\ \cline{2-10}
 $(0,0.7)$&$10^{-3}$&$0.0010 $     &0.064      &0.21      & 0.41     &0.56      &0.70      &0.82      &0.91 \\ \cline{2-10}
 &$10^{-5}$&$1.0\times 10^{-5}$      &0.0064      &0.044      &0.13      &0.22      &0.32      &0.42      &0.51 \\ \hline
 &$10^{-1}$&0.38      & 1.4   & 1.1  &   1.1  &   1.3   &  1.5    &    1.6  & 1.7 \\ \cline{2-10}
 $(1,0.3)$&$10^{-3}$&$0.0038 $     &0.14      &0.23      & 0.36     &0.53      &0.70      &0.84      &0.95 \\ \cline{2-10}
 &$10^{-5}$&$3.8\times 10^{-5}$      &$0.014$      &0.049      &0.11      &0.21      &0.33      &0.44      &0.53 \\ \hline

 &$10^{-1}$&0.56      &   1.2   &    1.6  &   2.1   &   2.5   &  2.8    &    3.2  & 3.0 \\ \cline{2-10}
 $(0,0.3)$&$10^{-3}$&$0.0056 $     &0.12      &0.34      & 0.66     &1.0      &1.3      &1.6     &1.7 \\ \cline{2-10}
 &$10^{-5}$&$5.6\times 10^{-5}$      &$0.012$      &0.074      &0.21      &0.40      &0.60      &0.85      &0.94 \\ \hline
\end{tabular}
\caption{Values of $\ep_{\mathrm{max}}$ that satisfies 
the required accuracy of $\Delta$
for each pair of 
$(B, \sigma)$ and different orders of asymptotic expansion, $N$. 
%When asymptotic expansion of higher-order is used, 
%$\ep_{\mathrm{max}}$ is larger, so that the analytic solution constructed is 
%sufficiently accurate until a relatively later moment in time. 
%For instance, one can see from the table that 
%when asymptotic expansion of first order is used, 
%numerical calculation 
%has to be started at $\ep=0.0064$ in order to maintain the accuracy 
%of order $10^{-5}$, for the profile with $(B,\sigma)=(0,0.7)$. 
%On the other hand, 
%if we use asymptotic expansion of seventh order, 
%we can follow time evolution of perturbation until 
%$\ep=0.51$, maintaining the accuracy of $10^{-5}$, 
%for the profile with $(B,\sigma)=(0,0.7)$.
%For instance, one can see from the table that the error of the analytic solution
%calculated at $\ep=0.94$ using asymptotic expansion of eighth order is expected to be of order $10^{-5}$ for curvature
%profile with $(B,\sigma)=(0,0.3)$.
}
\label{table}
%\end{center}
\end{table}

%%%%%%%%%%%%%%%%%%%%%%%%%%%%%%%%%%%%%%%%%%%%
\section{Discussion and conclusion}
\begin{figure}[h]
\begin{center}
\includegraphics[width=17cm,keepaspectratio,clip]{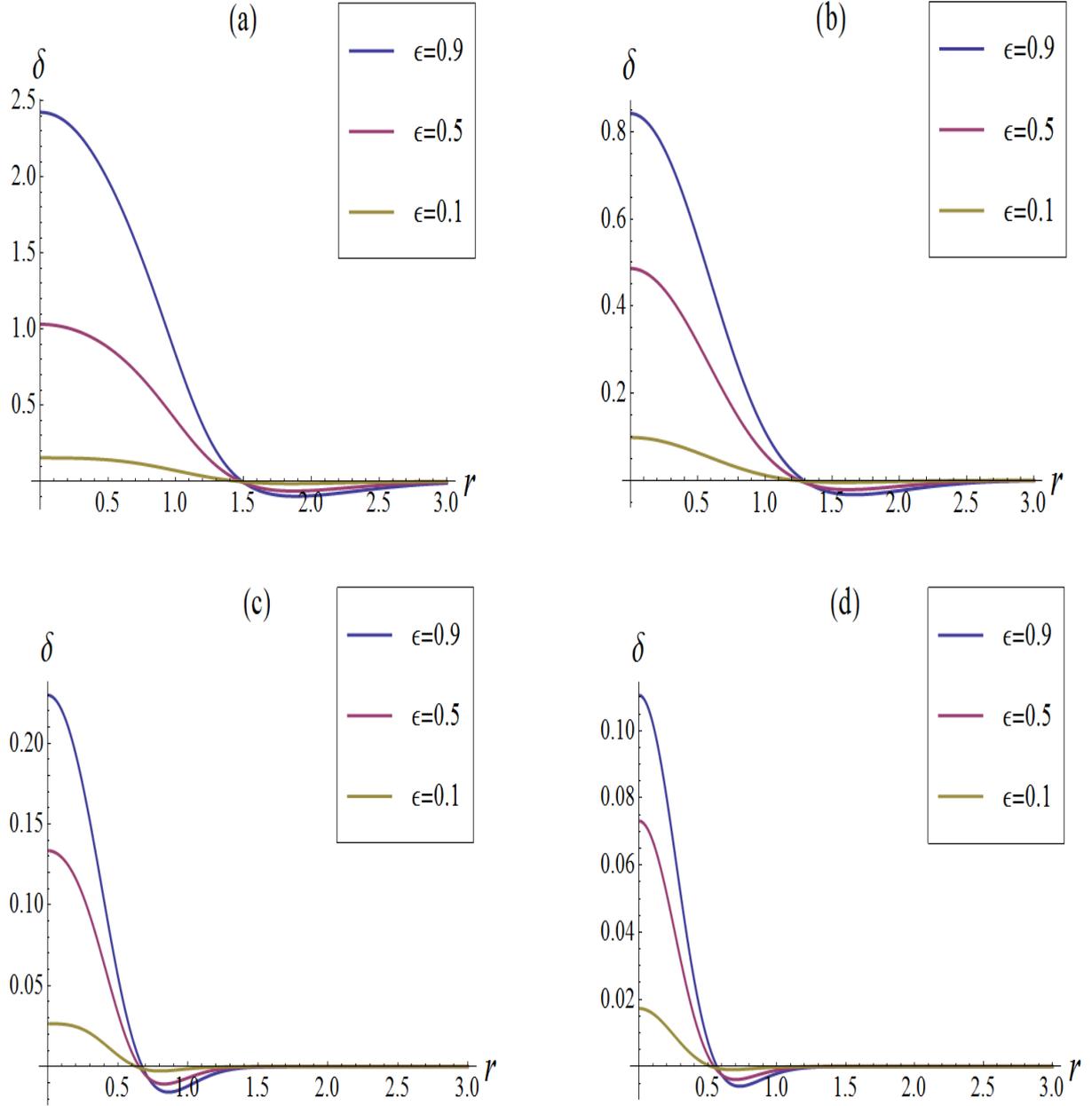}
\end{center}
\caption{An illustration of ``time evolution'' of the density perturbation profiles.
These were calculated when $\ep=0.1$, 0.5 and 0.9.
(a), (b), (c) and (d) were obtained from $(B,\sigma)=(1,0.7),(0,0.7),(1,0.3)$, and $(0,0.3)$, respectively.
One can see that the region with $\delta>0$, which corresponds to the central overdense region, is surrounded by the underdense region with $\delta<0$.
}
\label{overdensities}
\end{figure}

%%%%%%%%%%%%%%%%%%%%%%%%%%%%%%%%%%%%%%%%%%
In the present paper we have formulated a recursive method of
quasi-linearization which can yield appropriate initial condition 
for PBH formation consistent with general relativity. 
The evolution of the profiles of the energy density perturbation $\delta$ are
shown in Figure \ref{overdensities}. These profiles are calculated at
 $\ep=0.1,0.5$ and $0.9$.
One can see that the region with $\delta>0$, which corresponds to the central overdense region, 
is surrounded by the underdense region with $\delta<0$ so that (\ref{condfordelta}) is satisfied.
%This feature can be understood by noting that $\delta$ satisfies the following equation,
%\be
%\int^\infty_04\pi\delta R^2dR=0,
%\ee
%which results from the assumption that our solution coincides exactly with the background Friedmann solution at spatial infinity.

%We also introduce the total overdensity, denoted by $\bar{\delta}$ and defined as the proper volume integral of the 
%energy density perturbation over the overdense region:
We also introduce the averaged overdensity, denoted by $\bar{\delta}$
and defined as the energy density perturbation averaged over the
overdense region as follows:
\be
\bar{\delta}(t)\equiv\left(\fr{4}{3}\pi R(t,r_\mathrm{od})^3\right)^{-1}\int^{R(t,r_\mathrm{od})}_04\pi\delta R^2dR.
\ee
Here $r_\mathrm{od}(t)$ represents the comoving radius of the 
overdense region, which is numerically calculated from the solution of $\delta(t,r_{\mathrm{od}})=0$.
It turns out that $r_\mathrm{od}(t)$ is very close to $r_\mathrm{i}$
calculated from (\ref{rieq}), i.e. lowest-order expansion.
This feature can be directly observed in Figure 6, where the coordinate
with
$\delta=0$ hardly changes.

\begin{figure}[h]
\begin{center}
\includegraphics[width=14cm,keepaspectratio,clip]{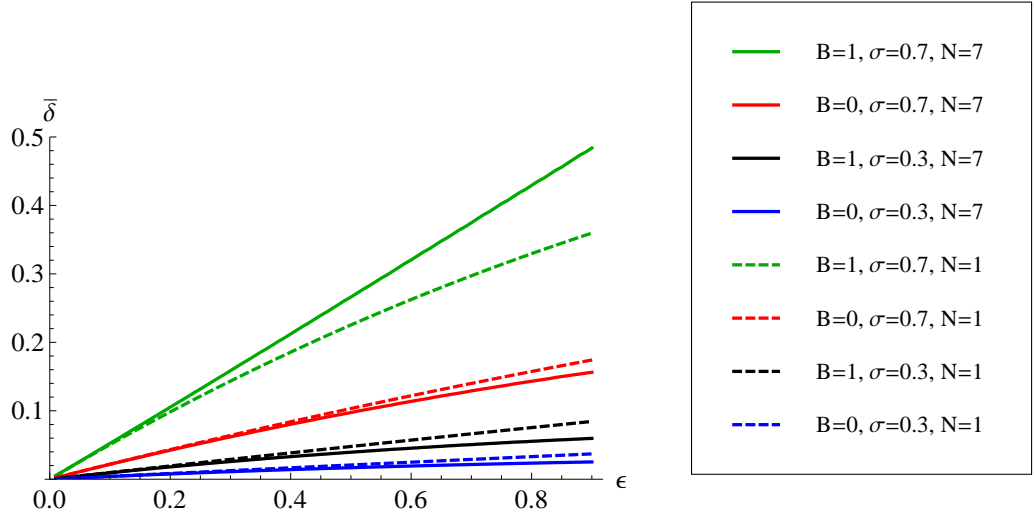}
\end{center}
\caption{Time evolution of the averaged overdensity $\bar{\delta}$ for each 
of the initial curvature profiles.
For comparison, the results obtained from the 
first-order asymptotic expansion are also shown by the dashed lines.
%One can confirm that higher-order corrections become more 
%and more important when $\epsilon$ gets closer to unity.
%When amplitude of initial curvature fluctuation is wider 
%and its profile is steeper, density perturbation in the central region
%becomes larger, so that
%$\bar{\delta}$ tends to be larger. Therefore, it is more 
%likely that wider and steeper initial curvature profiles lead to
%PBH formations after the perturbed region reenters the horizon.
}
\label{evolution-totaloverdensity}
\end{figure}

\begin{figure}[h]
\begin{center}
\includegraphics[width=14cm,keepaspectratio,clip]{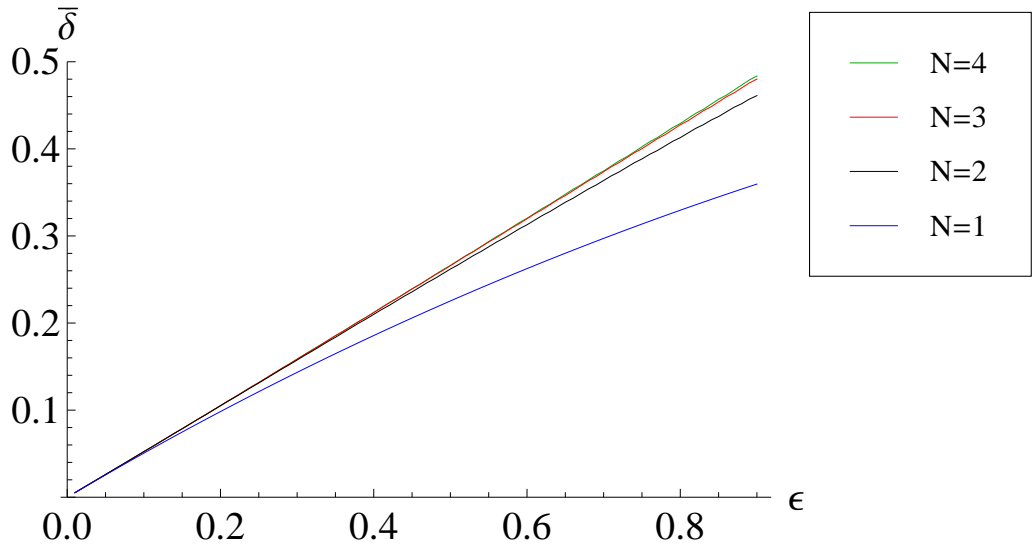}
\end{center}
\caption{Comparison of the time evolution of the averaged overdensity 
$\bar{\delta}$ for $N=1-4$.
The profile with $(B,\sigma)=(1,0.7)$ was used for these plots. 
%One can confirm that when $\ep\ll 1$, the plots coincide well with each other, but 
%as $\ep$ becomes larger, calculations using lower-order expansions start to deviate from those using higher-order ones. 
}
\label{deltabar_different_orders}
\end{figure}

The time evolution of the averaged overdensity $\bar{\delta}$ is shown 
in Figure \ref{evolution-totaloverdensity}.
For comparison, the results obtained using asymptotic expansions
 of first order are also shown.
From the plots, one can confirm that higher-order corrections 
become more important as $\epsilon$ gets closer to unity.
When the amplitude of initial curvature fluctuation is wider and 
its profile steeper, the density perturbation in the central region
becomes larger, so that
$\bar{\delta}$ tends to be larger. Therefore, it is more likely 
that wider and steeper initial curvature profiles lead to
PBH formation after the perturbed region reenters the horizon.
This confirms that considering the 
shape of profiles is crucial in the analysis of PBH formation. 
In addition, comparison of the time evolution of averaged overdensity
$\bar{\delta}$ 
for $N=1-4$ is shown 
in Figure \ref{deltabar_different_orders}.
The profile with $(B,\sigma)=(1,0.7)$ was used for these plots. 
When $\ep\ll1$, the plots coincide well with each 
other, but 
as $\ep$ becomes larger, calculations using lower-order expansions 
start to deviate from those using higher-order ones. 

We have analyzed various configurations of curvature perturbations
under the assumption of spherical symmetry to set up the initial
condition for the numerical analysis of 
PBH formation in an optimal way with the help of the
asymptotic expansion. 
In our analysis the  curvature profile
has a characteristic scale much larger than the Hubble radius initially, 
in accordance with the inflationary cosmology 
\cite{Sato:1980yn,Guth:1980zm,Starobinsky-1980} which predicts
formation of superhorizon-scale curvature perturbations
\cite{Mukhanov:1982nu,Guth:1982ec,Hawking:1982cz,Starobinsky:1982ee}.
This includes those perturbations which could lead to PBH formation
 \cite{PhysRevD.54.6040,
PhysRevD.42.3329,PhysRevD.50.7173,yokoyama-1997-673,
PhysRevD.58.083510,Jun'ichi1998133,kawasaki-1999-59,
PTPS.136.338,1475-7516-2008-06-024,PhysRevD.59.103505,
PhysRevD.63.123503,
PhysRevD.64.021301,Kawasaki:2007zz,Kawaguchi:2007fz}.
In a future paper we plan to calculate the probability of
realization of the curvature profiles discussed above.  Then 
we will eventually be able to relate the mass spectrum of PBHs
with the parameters of inflationary models.

\section*{Acknowledgments}

AGP acknowledges RESCEU for hospitality where this work was 
started.  
This work was 
supported in part by JSPS Grant-in-Aid for Scientific Research
No.\ 23340058 (JY), Grant-in-Aid for Scientific Research on 
Innovative Areas No.\  21111006 (JY), 
and Global COE Program ``the Physical Sciences Frontier'', MEXT, Japan.
This work has also benefited from exchange visits supported by a Royal
Society and JSPS bilateral grant.

\clearpage
\bibliographystyle{apsrev}
\bibliography{pbhdraft}

\end{document}